\documentclass[aps, amsmath,amssymb,pra,twocolumn,superscriptaddress]{revtex4-2}

\usepackage{graphicx}
\usepackage{float}
\usepackage{physics}

\usepackage{dcolumn}
\usepackage{bm}
\usepackage{tikz}
\usepackage{epstopdf}
\usetikzlibrary{decorations.pathreplacing}

\begin{document}

\title{Persistent current oscillations in a double-ring quantum gas}

\author{T. Bland}
\affiliation{Joint Quantum Centre (JQC) Durham-Newcastle, School of Mathematics, Statistics and Physics, Newcastle University, Newcastle upon Tyne, NE1 7RU, United Kingdom}
\affiliation{Institut f\"{u}r Experimentalphysik, Universit\"{a}t Innsbruck, Austria}
\author{I.\,V. Yatsuta}
\affiliation{Department of Physics, Taras Shevchenko National University of Kyiv, 64/13, Volodymyrska Street, Kyiv 01601, Ukraine}
\affiliation{Department of Particle Physics and Astrophysics, Weizmann Institute of Science, Rehovot 7610001, Israel}
\author{M. Edwards}
\affiliation{Department of Physics, Georgia Southern University, Statesboro, Georgia 30460-8031, USA}
\author{Y.\,O. Nikolaieva}
\affiliation{Department of Physics, Taras Shevchenko National University of Kyiv, 64/13, Volodymyrska Street, Kyiv 01601, Ukraine}
\author{A.\,O. Oliinyk}
\affiliation{Department of Physics, Taras Shevchenko National University of Kyiv, 64/13, Volodymyrska Street, Kyiv 01601, Ukraine}
\author{A.\,I. Yakimenko}
\affiliation{Department of Physics, Taras Shevchenko National University of Kyiv,
64/13, Volodymyrska Street, Kyiv 01601, Ukraine}
\author{N.\,P. Proukakis}
\affiliation{Joint Quantum Centre (JQC) Durham-Newcastle, School of Mathematics, Statistics and Physics, Newcastle University, Newcastle upon Tyne, NE1 7RU, United Kingdom}

\date{\today}

\begin{abstract}
Vorticity in closed quantum fluid circuits is known to arise in the form of persistent currents.
In this work, we develop a method to engineer transport of the quantized vorticity between density-coupled ring-shaped atomic Bose-Einstein condensates in experimentally accessible regimes. Introducing a tunable weak link between the rings, we observe and characterize the controllable periodic transfer of the current, and investigate the role of temperature on suppressing these oscillations via a range of complementary state-of-the-art numerical methods. Our set-up paves the way for precision measurements of local acceleration and rotation.
\end{abstract}

\maketitle

\section{Introduction}

Persistent currents---quantized flow of atomic Bose-Einstein Condensates (BECs) in closed circuits---enable fundamental studies of superfluidity and may lead to applications in high precision metrology and atomtronics \cite{amico2017focus,amico2021roadmap,amico2021atomtronic}, such as the recently observed atomtronic SQUID \cite{ryu2020quantum}.
The question of the generation and stability of the atomic persistent currents---which in the absence of external driving should be topologically protected---is of fundamental importance; thus it has been the subject of numerous experiments \cite{wright2000toroidal,ryu2007observation,ramanathan2011superflow,moulder2012quantized,ryu2013experimental,Wright2013,murray2013probing,beattie2013persistent,corman2014quench,aidelsburger2017relaxation,eckel2014hysteresis,eckel2016contact,cai2022persistent,del2022imprinting}.

Beyond the intriguing physics afforded by single ring systems, several proposals for two parallel or stacked rings have shown that the persistent current can tunnel between rings in both many-body \cite{richaud2017quantum,escriva2021static} and mean-field studies \cite{oliinyk2019tunneling,oliinyk2019symmetry,oliinyk2020nonlinear}. At the single-particle level, persistent current tunneling has been predicted between arrays of adjacent rings and in similar configurations \cite{polo2016geometrically,pelegri2017single,pelegri2019topological,pelegri2019topological2,perez2022coherent}. However, tunneling has been shown to be forbidden at the mean-field level \cite{bland2020persistent,perez2022coherent}. Such a set-up is the cold-atom analogue of a qubit made from adjacent superconducting loops, known as the Mooij-Harmans qubit \cite{mooij2005phase,mooij2006superconducting,astafiev2012coherent}, with control over the tunneling of persistent currents in a double-ring geometry emulating the original theoretical scheme.

The geometry considered in this work is of a BEC state in a co-planar, side-by-side, double-ring geometry \cite{bland2020persistent}, including a tunable weak link across the interface between the two rings, acting as a mediator for coherent transport of persistent current between them. Recently, a similar scenario was studied at the few-particle many-body level, and quantum phase-slips were observed in a double-ring lattice with a tunable weak link \cite{perez2022coherent}. At the many-body level, the coupling between adjacent rings can lead to superposition and entanglement of persistent current states, a feature not afforded at a mean-field description. However, at the mean-field level, one has a many-particle state that is highly controllable and robust to thermal and quantum fluctuations. In this scenario, one can either utilize a connecting secondary ring as a non-destructive measurement device for an experiment conducted in the primary ring, or as a unique double-ring experiment, where the dependency of external factors on the transfer of persistent current can act as a highly sensitive precision measurement device.

\begin{figure*}
\centering
\includegraphics[width=\textwidth]{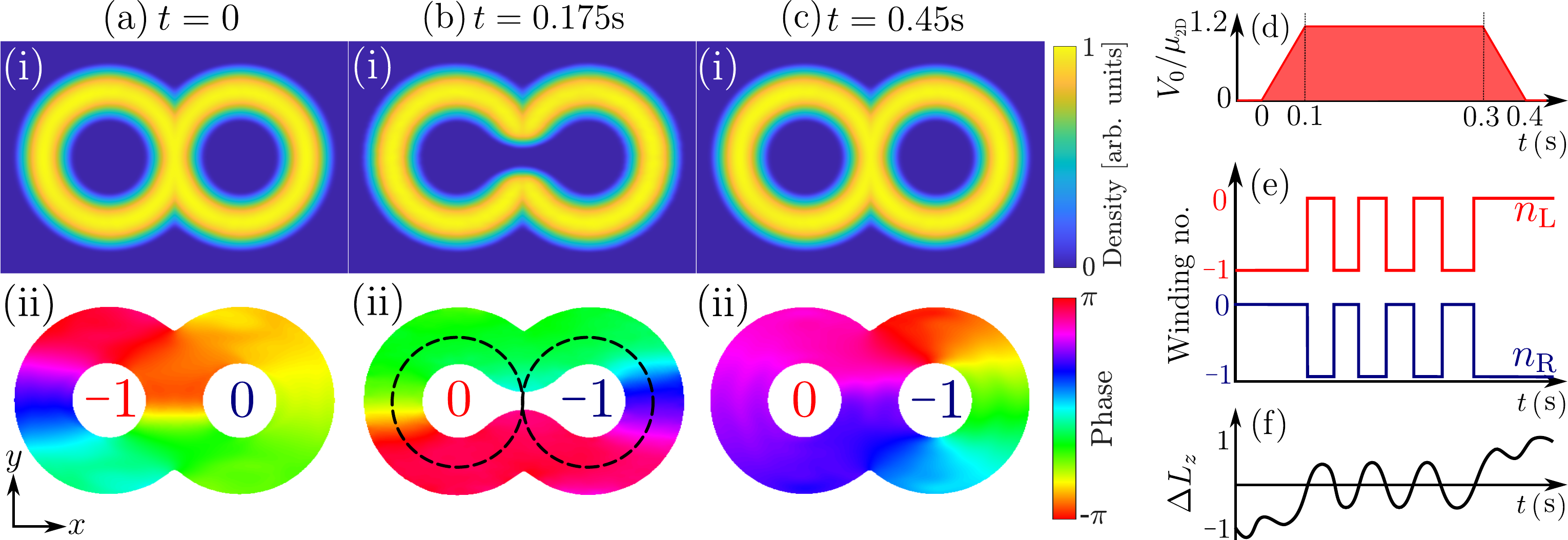}
\caption{Persistent current oscillations. (a) Double-ring ground state solution showing (i) density and (ii) phase. The phase shows an initial anti-vortex (clockwise circulation) imprinted into the left ring. (b) After opening the gate between two rings, the anti-vortex can periodically transfer. The persistent current is measured dynamically through the phase around the dashed ellipses in (ii), with current plot illustrating first instance of persistent current transfer to right ring. (c) The timing of closing the gate sets the final position of the persistent current. (d) Barrier amplitude as a function of time (in units of the system chemical potential). (e) Dynamic persistent current for the left $n_\text{L}$ (red) and right $n_\text{R}$ (blue) rings. (f) Difference in angular momentum between rings, see Eq.~\eqref{eqn:Lz}.}
\label{fig:1}
\end{figure*}

In this work, we address the dynamics of a single persistent current, controllably initiated in one of two density coupled rings, manipulated by an external barrier potential acting as a tunable weak link separating and re-connecting the two rings, as shown in Fig.~\ref{fig:1}(a)-(c) [top].
%
Such fundamental underlying physics is analyzed in detail theoretically and numerically for experimentally-relevant parameters in the quasi-two-dimensional (quasi-2D) limit, both 
in the context of the $T=0$ mean-field Gross-Pitaevskii equation and its appropriate dissipative and finite-temperature realizations.

In the pure ($T=0$) mean-field limit, we find the persistent current executes undamped oscillation cycles across the density threshold between the two rings, at a fixed frequency depending only on the geometry and the barrier amplitude. After introducing our double-ring geometry (Sec.~II), we first analyze such dynamics in the pure 2D limit,
with further insight into vortex dynamics in this limit provided by a semi-analytic toy quantum vortex kinetic model (Sec.~III).
The effect of thermal dissipation is then investigated (Sec.~IV) by the addition of a phenomenological damping term through full numerical simulations, revealing the damping of such oscillations, with our quantum vortex equation of motion providing a useful estimate of the thermal cloud dissipation rate required to halt the oscillations.
Our dissipative Gross-Pitaevskii model is further extended by the addition of fluctuations through the stochastic Gross-Pitaevskii model, revealing the robust existence of such oscillations even in the presence of (thermal) fluctuations.

To account for the self-consistent coupling of the condensate to a dynamical thermal cloud while simultaneously taking account of the full 3D geometry of our set-up, we also consider (Sec.~V) a generalized kinetic model, in which the self-consistently coupled thermal cloud--which modifies the effective trap felt by the condensate--is itself described by a collisionless Boltzmann equation.

Confirmation of the existence of persistent current oscillations across all studied models, including the full 3D kinetic model, offers further support towards the feasibility of experimental observation of such oscillations.

\section{Double-Ring Geometry}

The geometry considered throughout this work is that of two co-planar, partly overlapping density rings, whose planar 2D geometry \cite{bland2020persistent} is shown in Fig~\ref{fig:1} [top].

The double ring geometry is fixed by
\begin{eqnarray}
  V_\mathrm{ext}(\bm{\rho},z,t) &=& V_\mathrm{ext}(\bm{\rho},t) + V_\perp(z)  \nonumber \\ &=& V_\text{rings}(\bm{\rho})+V_\text{barrier}(\bm{\rho},t) +V_\perp(z).
  \label{eqn:Vext}
\end{eqnarray}
This is the superposition of a potential of two 2D conjoined rings
\begin{align}\label{eq:Vd2}
     V_\text{rings}(\bm{\rho})=\frac{1}{2}m\omega_\rho^2\min\left(({\rho_{-}}\hspace{-0.8mm}-R)^2, (\rho_+\hspace{-0.8mm}-R)^2\right)\,,
\end{align}
with a controllable time-dependent barrier $V_\text{barrier}(\bm{\rho},t)$, and
a harmonic transverse confining potential
\begin{align}
    V_\perp(z) = \frac{1}{2} m\omega_z^2 z^2 \,.
\end{align}
The parameters of each ring have been chosen to match the single-ring experiment of Ref.~\cite{Wright2013} based on $^{23}$Na atoms, with an {\it s}-wave scattering length $a_s=2.75\,$nm, but with a different fixed total atom number of $N=10^6$. 
As such, we have chosen ring radii $R=22.6\,\mu$m, $\rho_\pm = \sqrt{(x\pm R)^2 + y^2}$, a radial trapping frequency $\omega_\rho = 2\pi\times134\,$Hz, and a tighter transversal trapping frequency $\omega_z = 2\pi\times550\,$Hz, such that the system is in the quasi-2D regime, with dominant dynamical features arising within the central plane (for info, our parameters give $\mu/\hbar \omega_z < 4$).
The barrier potential controlling the degree of connectedness of the two rings has the form
\begin{align}\label{eq:Vb}
V_\text{barrier}(\bm{\rho},t)=V_0(t)\Theta(R-|x|)e^{-y^2/2\sigma^2}\,,
\end{align}
featuring a time-dependent amplitude $V_0(t)$, whose maximum value is slightly above the system chemical potential $\mu$, and a barrier width $\sigma = 3.44\,\mu$m, with $\Theta(x)$ denoting the Heaviside (step) function.


\section{Persistent Current Oscillations at Mean-Field (Gross-Pitaevskii) level}\label{sec:2}
It has been shown previously that persistent currents in a co-planar double-ring quantum gas will not tunnel between rings \cite{bland2020persistent}. In this section, we show that including a barrier potential between the rings does in fact facilitate such persistent current transfer in the pure $T=0$ condensate limit.

We assume here tight transverse confinement, such that the full condensate wave function can be expressed in the form 
$\Psi({\bf r},t) = \psi(\bm{\rho},t)\equiv\psi(x,y,t) \phi(z)$, where $\phi(z)$ denotes a transverse Gaussian of width $l_z = \sqrt{\hbar / m \omega_z}$.
This allows us to
start our theoretical investigations in the context of the pure two-dimensional (2D) 
%
%
Gross-Pitaevskii equation (GPE) describing the evolution of the  2D wave function $\psi(\bm{\rho},t)\equiv\psi(x,y,t)$ via
\begin{align}
i\hbar \frac{\partial}{\partial t}\psi(\bm{\rho},t) = \bigg[\hat{\mathcal{H}}_\text{GP}[\psi]-\mu_\text{2D}\bigg]\psi(\bm{\rho},t)\,,
\label{eqn:GPE}
\end{align}
where
\begin{align}
\hat{\mathcal{H}}_\text{GP}[\psi]=-\frac{\hbar^2}{2m}\nabla^2+V_\text{ext}(\bm{\rho},t)+g_\text{2D}|\psi(\bm{\rho},t)|^2\,,
\label{eqn:HGP}
\end{align}
is the 2D Gross-Pitaevskii operator, and the density is normalized such that $\iint \text{d}x\text{d}y\,|\psi(\bm{\rho},t)|^2=N$. Here, ${g_\text{2D}=g/\sqrt{2 \pi} l_z = \sqrt{8\pi}\hbar^2a_s/ml_z}$ is the effective two-dimensional two-body interaction coupling ($g$ denotes the 3D coupling), with associated {\it s}-wave interaction strength $a_s=2.75\,$nm, $\mu_\text{2D}$ is the 2D chemical potential (related to the full 3D chemical potential $\mu$), $m$ the $^{23}$Na atomic mass.
For our chosen parameters,
$\mu_\text{2D}=14.93\hbar\omega_\rho$ before the addition of the potential (i.e.~for $V_0=0$). All simulations are performed in a $L_x\times L_y = 120\times80\mu$m$^{2}$ grid, with $768\times512$ grid points.

The barrier amplitude is increased/decreased linearly, with its maximum constant value slightly exceeding the chemical potential, in order to establish an effective barrier.
Given the physical importance of the relation of the barrier amplitude to the chemical potential, throughout this work, $V_0(t)$ will always be scaled to the relevant chemical potential [see Fig.~\ref{fig:1}(d)].


To analyze the system dynamics, we first numerically obtain the ground state of the system in the absence of the barrier (i.e.~when $V_0=0$) via imaginary time propagation. 
We then initialize our persistent current oscillator by phase imprinting a $2\pi$ clockwise winding around the center of the left ring,
such that the system state can be characterized by  the corresponding winding numbers $n_\text{L}=-1$ (left ring) and $n_\text{R}=0$ (right ring, no persistent current). The persistent current, or winding number, is akin to a ``ghost" vortex in the center of the annulus, and this language will be used interchangeably (with winding number of $+1$ ($-1$) respectively corresponding to the presence of a ``ghost'' vortex (anti-vortex). After phase imprinting, we allow for 100ms of thermalization, leading to the initial condition shown in Fig.~\ref{fig:1}(a). Previous work in this geometry has shown that the ghost vortex (or anti-vortex) remains trapped in its initial state \cite{bland2020persistent}. However, the inclusion of an external barrier potential changes the system topology from a 2-torus to a torus, allowing for the transfer of the current between rings.

\subsection{Undamped oscillations}

After an initial thermalization period, the barrier potential is linearly ramped up to the maximum $V_0/\mu_{\text{2D}} = 1.2$ [Fig.~\ref{fig:1}(d)], with such opening of the gate (around the $(x,y)=(0,0)$ region) allowing for vortex transfer between the two rings. Once $V_0$ exceeds $\mu$ the persistent current begins to oscillate between the two rings [Fig.~\ref{fig:1}(e)]. In this set of simulations, the gate is held open for 200 ms and the current oscillates at a fixed period of $\sim$50 ms. Then the potential amplitude is linearly reduced, and the position of the vortex as $V_0$ crosses $\mu$ sets in which of the two rings the persistent current will reside and be detectable.
In the example shown, the persistent current has transferred from its initial state [left ring, Fig.~\ref{fig:1}(a)(ii)] to the right ring [Fig.~\ref{fig:1}(c)(ii)]. Note that, accompanying the persistent current oscillation, is a small transfer of atom number, as measured either side of $x=0$, with $<$0.2\% of the total atom number transferring with the persistent current.
Although insignificant in terms of atom numbers, such a transfer does nonetheless have a numerically measurable impact on the angular momentum dynamics, to be described below.

\begin{figure}
    \centering
    \includegraphics[width=\columnwidth]{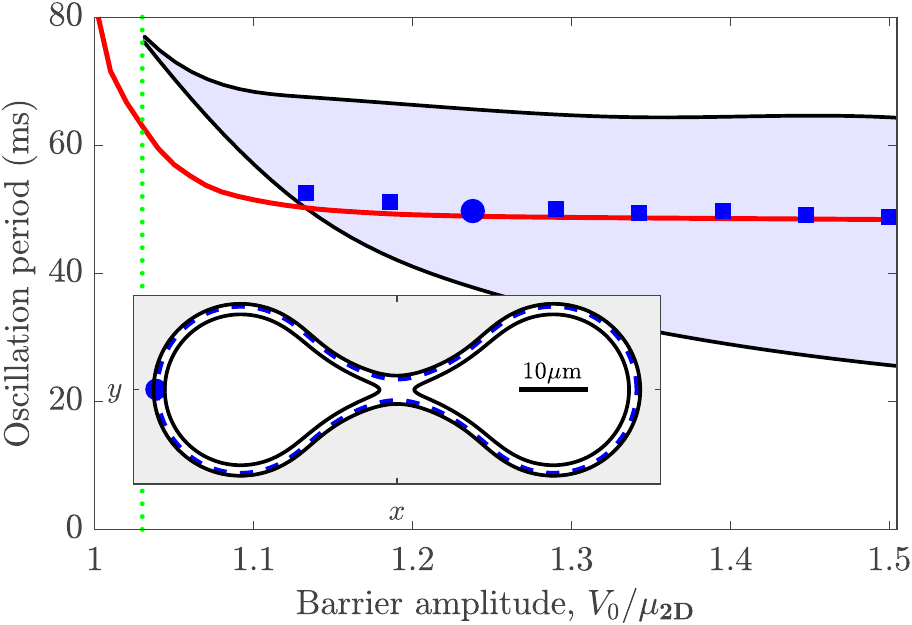}
    \caption{Persistent current oscillation period as a function of maximum barrier amplitude. Main plot: Oscillation period  obtained from the zero-temperature GPE (red curve). The blue markers are the equivalent data from the semi-analytic model (see main text). The initial vortex position for the analytic model is chosen to give the correct oscillation period for $V_0=1.24\mu_{\text{2D}}$ (blue circle), and then taken as an initial condition for square points. The filled region indicates the analytic oscillation period range. The vertical green dotted line corresponds to the point where $V_0$ is equal to the chemical potential {\em in the presence of} the barrier, thus denoting the parameter regime beyond which the toy model acquires its meaning. Inset: vortex trajectories in the double-ring system. Outer black line sets upper bound to the analytic model. For $V_0/\mu_{\text{2D}}<1$ the trajectories do not connect between rings, and there is no vortex transfer, as shown by disconnected lines inside, giving the lower bound to the filled region in the main plot. Dashed blue curve corresponds to the vortex orbit with $V_0/\mu_{\text{2D}}=1.24$.}
    \label{fig:2}
\end{figure}

It is important to note that the transfer is instantaneous when the vortex line crosses the threshold between rings. The winding is extracted dynamically via the azimuthal phase measured at a distance $R$ from the center of each ring [e.g.~dashed circles in Fig.~\ref{fig:1}(b)(ii)]. Therefore, if a vortex is contained anywhere within the circle of radius $R$ from the ring's center the phase measured azimuthally will wind by a factor of $2\pi$, otherwise it will return to its initial value. It is possible to further characterize the dynamic vortex position through measurement of the angular momentum in each ring, and define the angular momentum difference as $\Delta L_z = \langle L_{z,\text{L}}\rangle -  \langle L_{z,\text{R}} \rangle$, where
\begin{align}
    \langle L_{z,\text{\{L,R\}}}\rangle &= \frac{i\hbar}{N_\text{\{L,R\}}}\iint_{\mathcal{R}}\text{d}x\text{d}y\,\psi^\star\left(y\frac{\partial}{\partial x} - (x \pm R)\frac{\partial}{\partial y} \right)\psi
    \label{eqn:Lz}
\end{align}
with $N_{\text{L}/\text{R}}$ the particle number in the left/right ring, and the integration region $\mathcal{R}$ is over the left/right side of the box, accordingly.
From this equation, we associate a value of ${\Delta L_z\approx-1~(+1)}$ when an anti-vortex is centered in the left (right) ring, with intermediate values indicating a vortex between $x=(-R,R)$, where one could approximate the vortex position along the $x$-axis as $(\Delta L_z)R$. A similar measure was applied in Ref.~\cite{gallemi2015coherent} to determine the angular momentum difference between two components. An example trajectory of $\Delta L_z$ is shown in Fig.~\ref{fig:1}(f). Each crossing of $\Delta L_z=0$ corresponds to the transfer of persistent current. Whilst the barrier is open the amplitude of $\Delta L_z$ does not exceed 0.5, suggesting that the vortex does not travel near to the center of a ring unless the barrier is closed.


Next, we investigate the oscillation period of the persistent current as a function of the maximum barrier amplitude, still in the zero-temperature limit [Fig.~\ref{fig:2}]. If the barrier amplitude is smaller than the chemical potential then there are no observed oscillations, and the vortex remains in its starting ring indefinitely. However, for all $V_0>\mu_{\text{2D}}$ the vortex exhibits symmetric oscillations about the center of the system. For larger barrier amplitudes ($V_0>1.2\mu_{\text{2D}}$) the resulting oscillation period is almost constant, at around $50$ ms. We have explored varying the linear ramp gradient, and find this only has a weak effect on the oscillation period (typically $\pm1$ ms), however if the barrier amplitude gradient is large, reaching its maximum value in $<20$ ms this strongly perturbs the system, injecting vortex pairs and high-amplitude noise.

We can gain an insight into the vortex dynamics through a toy kinetic model.
The velocity of a quantum vortex in an inhomogeneous condensate can be written as \cite{groszek2018motion}:
\begin{align}
 \textbf{v} = \frac{\hbar}{m}\left(\nabla \Phi - \hat{\bm{\kappa}} \times \nabla \ln{\sqrt{n}}\right)\,,
 \label{eqn:vortex}
\end{align}
where $\hat{\bm{\kappa}} = \kappa s \cdot \hat{\textbf{e}}_{z}$, the integer $s$ is the vortex winding number, ${\kappa = h/m}$ is the quantum of circulation, and $n = |\psi|^2$ and $\Phi$ are the density and phase of the condensate in the absence of a vortex, respectively. We take the numerically obtained stationary solution with fixed $V_0$ and choose an appropriate starting position for the quantum anti-vortex near to the point of the lowest density on the inside edge of the annulus. Using such initial condition, we iteratively solve Eq.~\eqref{eqn:vortex} to follow the vortex trajectory as it traverses around the hourglass-shaped inside edge. If the orbit is connected then the vortex traverses both rings, and the time to return to the initial position gives the semi-analytic oscillation period; otherwise the orbit is disconnected, and the vortex does not transfer. The vortex height on the density distribution is constant, and in order for this to vary we need to introduce dissipation, which will be the topic of the next section.

The range of semi-analytically obtained oscillation periods are contained within the shaded region of Fig.~\ref{fig:2}, bounded by two lines obtained as described below.
The upper boundary is given by initializing the vortex position at the `inner' Thomas-Fermi radius (i.e.~closer to the centre, $\rho<R$, where the barrier $V$ is greater than the chemical potential); this point has been chosen in order to preserve the ``ghost'' vortex identity, because, if the vortex were any closer to the peak density (at $\rho=R$) it would cease to be a ``ghost" vortex. 
The lower boundary is given by the smallest connected loop across both rings. The corresponding orbits are shown as solid lines on the inset to Fig.~\ref{fig:2}. 
In setting up such a toy model, we need to consider the densities in the {\em presence} of the separating barrier $V_0$. Due to our constraint on fixed atom number, this thus corresponds to a slightly higher chemical potential than the one before the addition of the barrier. As a result, the toy model can only give results beyond such point, indicated by the vertical dashed green line, at which $V_0 \sim 1.028 \mu_{\rm 2D}$.

Moreover, as such model ignores the role of self-consistent vortex-sound interactions on vortex dynamics (considered in detail in the related setting of Ref.~\cite{vortex-crosstalk}) our semi-analytical predictions cannot fully describe the dynamics in the regime when the barrier height is too close to the chemical potential, although such effects are of course fully captured within GPE numerical simulations.
%
%
%
With that in mind, we can nonetheless further test this model through direct comparison to the GPE results. First, we find the initial vortex position that corresponds to an oscillation period matching the GPE for a sufficiently high barrier height, arbitrarily chosen here as $V_0/\mu_{\text{2D}}=1.24$ [Fig.~\ref{fig:2} blue circle, and corresponding orbit in the inset]. 
Using exactly this initial vortex position for different $V_0$, we then  obtain the oscillation periods from within our model;
remarkably, we find all results to lie within the expected band and, for values sufficiently higher than the chemical potential (here $V_0/\mu_{\text{2D}}\gtrsim 1.13$, such that a connected orbit is still obtained), to give excellent agreement to the GPE, thus demonstrating the semi-quantitative validity of such an intuitive toy model.

\subsection{Observation of a beating effect}

\begin{figure}
    \centering
    \includegraphics[width=\columnwidth]{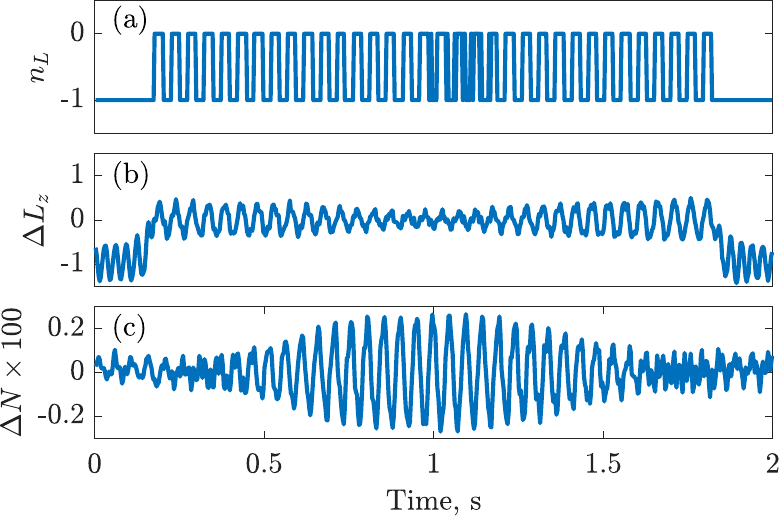}
    \caption{Beating effect at long times. $\gamma=0$ simulation extended to $t=2$s to show the collapse revival of $\Delta L_z$ oscillations. Shown are: (a) winding number oscillations in the left well, $n_\text{L}(t)$, (b) angular momentum difference $\Delta L_\text{z}(t)$, and (c) fractional population difference $\Delta N = (N_\text{L} -N_\text{R})/(N_\text{L} + N_\text{R})(t)$.}
    \label{fig:beating}
\end{figure}

The preparation of the initial state can cause small oscillations in the atom number between the two rings. The full impact of these oscillations is only observable in the long-time limit. We repeat the procedure of Fig.~\ref{fig:1},  but now monitoring the temporal evolution when the gate is kept open (at $V_0/\mu_{\text{2D}}=1.2$) for the even longer period of time of 1.6s. Figure~\ref{fig:beating} shows (a) the winding number oscillations in the left well, (b) the angular momentum difference, and (c) the fractional population difference $\Delta N = (N_\text{L} -N_\text{R})/(N_\text{L} + N_\text{R})$.
Analyzing the dominant frequencies through taking the respective discrete Fourier transforms of (b) and (c), we can infer a difference of $\sim 1.5$ms in their respective oscillation periods, which accounts for a beating period of $\sim 1.5$s, consistent with the features seen in the angular momentum difference.
We thus conclude that the observed beating (which appears to exhibit no noticeable damping in the $\gamma=0$ limit) is a result of coupling of the relative angular momentum to the relative fractional population difference. When the $\Delta L_z$ amplitude is at its smallest, this effect can produce glitches in our winding number measurements, as seen at $\sim1.1$s [Fig.~\ref{fig:3}(a)(ii)].
The wait time after the phase-imprinting protocol but prior to opening up the barrier controls the phase within a given beating cycle.
For the remainder of the work, we focus our analysis on timescales (broadly) consistent with a single beating half-cycle, in order to best highlight the role of dissipation and fluctuations.


\section{Finite Temperature effects:\\ Role of dissipation and fluctuations}

Actual experiments are typically conducted at low, but non-zero, temperatures---and so are prone to both dissipation and fluctuations, which we consider in this section. The net effect of such contributions, as described in detail below, is to damp, or even completely suppress, persistent--current oscillations. 

For a more stringent test of the mean-field predictions, we supplement our 2D model by additional contributions: In Sec.~IV A, we assess the role of dissipation through the addition of phenomenological damping to the GPE. We then further add corresponding thermal fluctuations to the model in Sec.~IV B. The generalization to full 3D considerations are then discussed in Sec.~V.

\subsection{Inclusion of Phenomenological Dissipation:\\ 
The Dissipative Gross-Pitaevskii Equation}

\begin{figure}
    \centering
    \includegraphics[width=1\columnwidth]{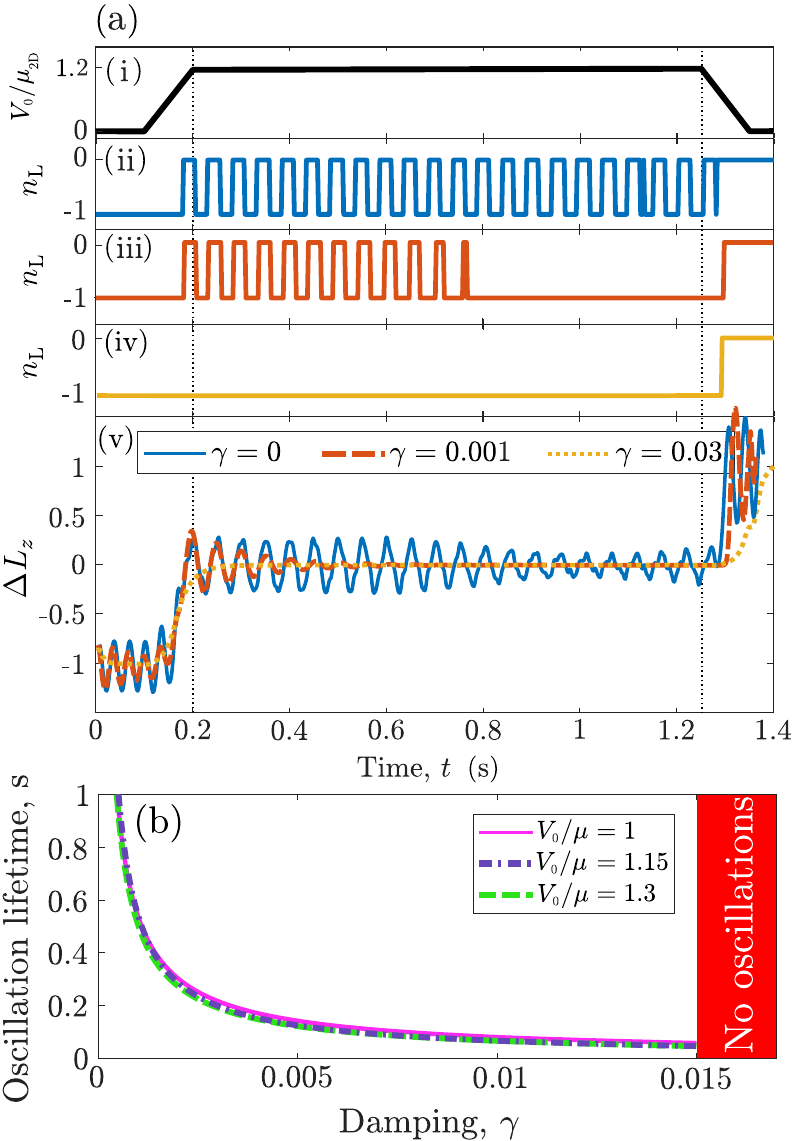}
    \caption{Suppression of persistent current oscillations. (a)(i) Barrier amplitude. Winding number oscillations for (ii) $\gamma = 0$, (iii) $\gamma = 0.001$, (iv) $\gamma = 0.03$. (v) Difference in angular momentum eigenvalues, $\Delta L_z = \langle L_{z,\text{L}}\rangle - \langle L_{z,\text{R}}\rangle$. (b) Oscillation lifetime as a function of maximum barrier amplitude and $\gamma$. There are no oscillations for $\gamma>\gamma_\text{cr}=0.015$.}
    \label{fig:3} 
\end{figure}

To include the dissipative effects of temperature, we extend our 2D model from Eq.~\eqref{eqn:GPE} to the damped (or dissipative) GPE, given by
\begin{align}
i\hbar \frac{\partial}{\partial t}\psi(\bm{\rho},t) = (1-i\gamma)\bigg[\hat{\mathcal{H}}_\text{GP}[\psi]-\mu_\text{2D}\bigg]\psi(\bm{\rho},t)\,.
\label{eqn:dGPE}
\end{align}
Here damping is phenomenologically included in the dimensionless parameter $\gamma \ll 1$,
with $\hat{\mathcal{H}}_\text{GP}$ unchanged from Eq.~\eqref{eqn:HGP}. 

As previously reported, in the zero temperature model ($\gamma=0$), oscillations continue indefinitely for as long as the barrier $V_0$ exceeds the system chemical potential. 

Having addressed the issue of the changing amplitude in $\Delta L_z$ in the $\gamma=0$ limit of a pure GPE,
we next investigate the role of damping on the lifetime of the oscillations. The results of our analysis using the dissipative GPE is shown, for $0 < \gamma \ll 1$, in Fig.~\ref{fig:3}. 

As evident from Fig.~\ref{fig:3},  the oscillations halt after 600ms for small $\gamma=0.001$ [panel (a)(iii)], as opposed to the undamped $\gamma=0$ oscillations shown in panel (a)(ii). Increasing to larger $\gamma=0.03$ the vortex does not transfer at all until (potentially) when the gate is closed [Fig.~\ref{fig:3}(a)(iv)]. Evaluation of $\Delta L_z$ for $\gamma = 0.03$ shows that the vortex becomes trapped at the center of the system [Fig.~\ref{fig:3}(a)(v)], and as the gate is closed the vortex is forced into the center of a randomly chosen ring, independent of the initial condition (i.e.~the vortex may stay in the same ring, or cross to the other ring -- the latter occurring in Fig.~\ref{fig:3}(a)). 

Intermediate values of $\gamma$ reveal the oscillation damping, with the $\Delta L_z$ oscillation amplitude rapidly decreasing to 0, an effect quite distinct to the $\gamma=0$ beating discussed above.

We also map out the oscillation lifetime, defined as the time from the barrier opening to the last vortex transfer, as a function of $\gamma$: our results are shown in Fig.~\ref{fig:3}(b). For small $\gamma<10^{-3}$ the oscillations are long-lived, surviving for $\sim1$s. The lifetime rapidly decreases with $\gamma$, and at $\gamma>\gamma_\text{cr}=0.015$ the lifetime is smaller than half of the oscillation period, trapping the persistent current in its initial ring. The lifetime of the oscillations is only weakly dependent on the maximum barrier amplitude. 

In order to obtain an estimate for $\gamma_\text{cr}$, we return to the vortex kinetic model. As in the damped GPE, dissipation in our model is included by replace the Hamiltonian from Ref.~\cite{groszek2018motion} with the dissipative one [$\hat{\mathcal{H}}_\text{GP}\to(1 - i\gamma)\hat{\mathcal{H}}_\text{GP}$] to give the next correction for the velocity of the vortex core
\begin{equation}
   \textbf{v} = \frac{\hbar}{m}\left(\nabla(\Phi - \gamma \ln{\sqrt{n}}) - \hat{\bm{\kappa}} \times \nabla( \ln{\sqrt{n}} + \gamma \Phi ) \right)\,,
   \label{eq:toymodel}
\end{equation}
Discarding terms with $\nabla \Phi$ (which are small compared to contributions from the density gradient) gives
\begin{equation}
   \textbf{v} = - \frac{\hbar}{m}\left(\gamma \nabla \ln{\sqrt{n}} + \hat{\bm{\kappa}} \times \nabla \ln{\sqrt{n}}\right) = \textbf{v}_{\text{r}} + \textbf{v}_{\phi}\,,  \label{eq:toymodelcomoving}
\end{equation}
where $\textbf{v}_{\text{r}}$ corresponds to radial vortex motion to larger or smaller densities, i.e.~changing the orbit length, and $\textbf{v}_{\phi}$ corresponds to the azimuthal vortex displacement along the orbit trajectory at constant density. We can get a useful expression if we compare small displacements of the vortex along above-mentioned directions through the ratio
\begin{align}
    \frac{|\textbf{v}_{\text{r}}|}{|\textbf{v}_{\phi}|} \equiv \frac{\text{d} l_{r}}{\text{d} l_{\phi}} = \frac{|\gamma \nabla \ln{\sqrt{n}} |}{ |\hat{\bm{\kappa}} \times \nabla \ln{\sqrt{n}}|} = \gamma\,.
\end{align}
We can use this to estimate the critical $\gamma$ at which the vortex drifts from an initially connected to a disconnected orbit. For this, we take the vortex initially placed at the blue point in the inset of Fig.~\ref{fig:2} (for which the correct oscillation period was obtained in the $\gamma=0$ limit) and find the smallest value of $\gamma$ at which the vortex meets the threshold between rings, but does not cross it. Here, the displacement along the radial direction is the distance from the Thomas-Fermi radius to the point with the shortest connected orbit, so $\text{d}l_r \approx\Delta l_r = 0.87\mu$m, and the azimuthal displacement is a quarter of the total length of the vortex orbit, $\text{d}l_\phi \approx\Delta l_\phi = 44.27\mu$m. For these values, we get $\gamma_{\text{cr}} \approx 0.0197$, which is reasonably close to the observed value $0.015$. As we can see, $\gamma_{\text{cr}}$ depends on the geometry of the system. This fact has some important consequences: different $\gamma_{\text{cr}}$ for different values of $V_{b}$ which we, however, do not observe from numerical investigation (see Fig. \ref{fig:3}); by making the radius of the ring smaller or by making the ring wider (shallower density gradient) we can increase $\gamma_{\text{cr}}$, thus making the oscillations more robust to finite temperature effects.

\begin{figure*}[!ht]
    \centering
    \includegraphics[width=2\columnwidth]{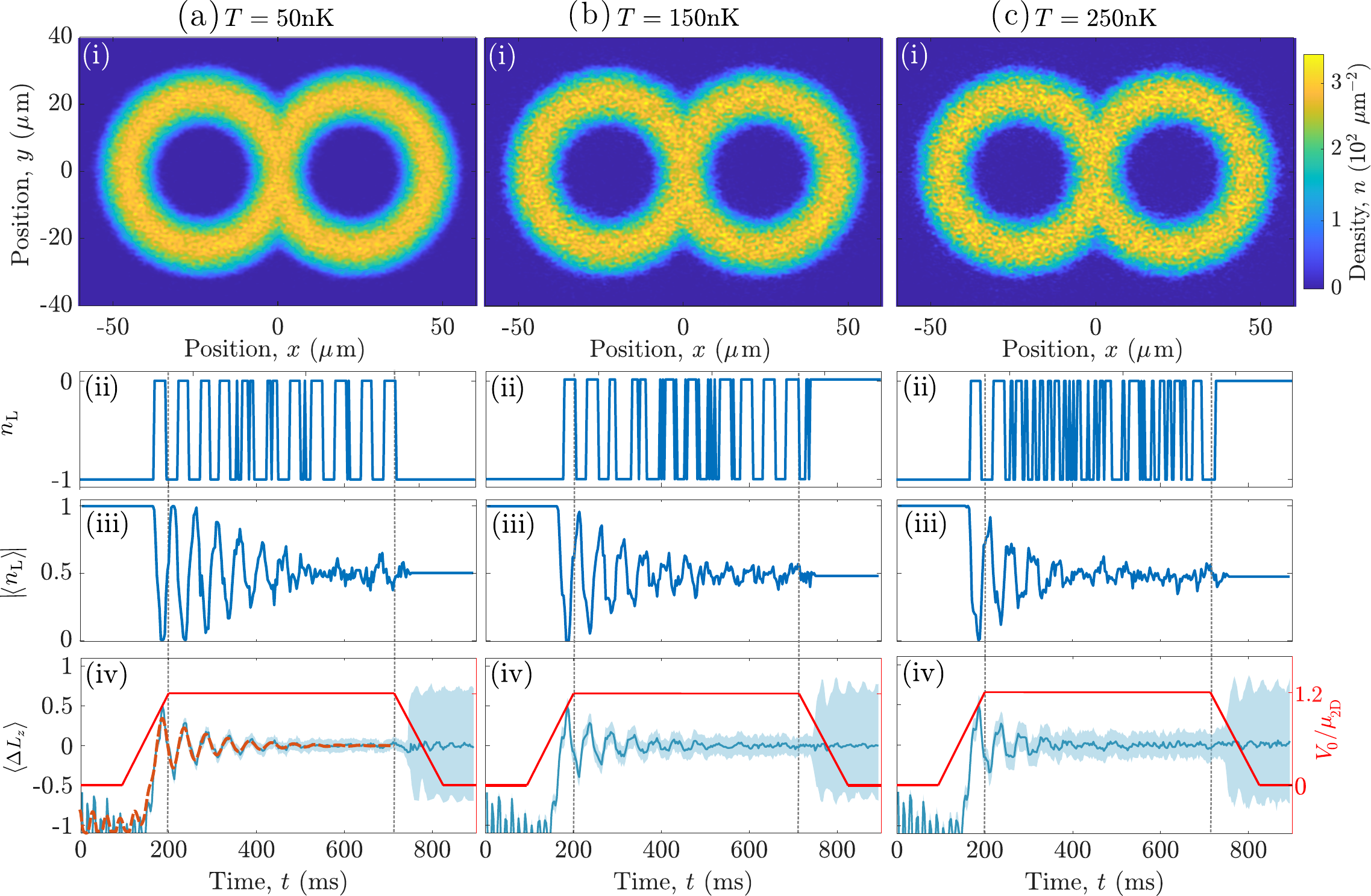}
    \caption{Persistent current oscillations for increasing temperature from left to right columns. Rows: (i) equilibrium SPGPE density profiles. (ii) Example persistent current oscillation in the left ring for a single run. (iii) winding number oscillations averaged over 100 runs. The absolute value $|\langle n_{\text{L}}\rangle|$ can be interpreted as a probability of finding the vortex in the left ring. (iv) Mean angular momentum difference between rings. Shading indicates one standard deviation from the mean (solid line). Red curve indicates barrier ramp protocol. Dashed line in (a)(iv) is the equivalent $T=0$, $\gamma=0.001$ data from Fig.~\ref{fig:3}.}
    \label{fig:4}
\end{figure*}

Having identified the dissipative persistent current oscillation feature, we next investigate the role of thermal fluctuations.

\subsection{Inclusion of fluctuations:\\ The (Projected) Stochastic Gross-Pitaevskii Model}

To include the effects of thermal fluctuations, we further extend our model in Eq.~\eqref{eqn:dGPE} to the Stochastic Projected Gross-Pitaevskii equation (SPGPE) \cite{Gardiner_2003,Bradley_2008,blakie2008dynamics,proukakis2008finite,gases2013finite}, along similar lines to previous studies in single ring-trap geometries~\cite{rooney2013persistent,gallucci2016engineering,Eckel_2018,szigeti21-superflowdecay}. In this formalism, the energy modes of the system are decomposed into the low-energy coherent region, described by the multi-mode order parameter $\Psi(\bm{\rho},t)$ mapping to the so-called `classical' or c-field' region, and the high-energy incoherent region which is assumed to play the role of a static heat bath of temperature $T$. Individual trajectories of the coherent region dynamics evolve according to the stochastic equation of motion \cite{proukakis2008finite}
\begin{align}
i\hbar \frac{\partial}{\partial t}\Psi = \hat{\mathcal{P}}\Bigg\{&(1-i\gamma)\bigg[\hat{\mathcal{H}}_\text{GP}[\Psi]-\mu_\text{2D}\bigg]\Psi+\eta(\bm{\rho},t)\Bigg\}\,,
\label{eqn:SPGPE}
\end{align}
describing their coupling to the higher-lying modes, where again $\hat{\mathcal{H}}_\text{GP}$ is unchanged from Eq.~\eqref{eqn:HGP}. The complex Gaussian noise satisfies $\expval{\eta^\star(\bm{\rho},t)\eta^\star(\bm{\rho}',t)}=\expval{\eta(\bm{\rho},t)\eta(\bm{\rho}',t)}=0$ and $\expval{\eta^\star(\bm{\rho},t)\eta(\bm{\rho}',t')}=2\gamma k_B T /\hbar\delta(\bm{\rho}-\bm{\rho}')\delta(t-t')$. The projector $\hat{\mathcal{P}}$ implements the energy cut-off, ensuring that the occupation of the largest included mode has average occupation of order unity. The energy cut-off here is fixed to $\epsilon_\text{cut}(\mu_{\text{2D}},T)=3\mu_{\text{2D}}$, consistent with previous studies \cite{blakie2008dynamics}, and the 2D chemical potential remains fixed to $\mu_\text{2D}=14.93\hbar\omega_\rho$. 

Each numerical realization has a different dynamical noise field, and can be {\em qualitatively} interpreted as a single experimental run (in the sense that an ensemble over numerical runs should produce the same results about mean values and fluctuations as an ensemble over many experiments).
The procedure is to simulate the dynamical setting multiple times based on different random noise sampling, and then extract appropriately-averaged physical quantities.

In each numerical realization for a given temperature, an initial state is generated by dynamical equilibration from a noisy initial field, leading to a state in thermal equilibrium with approximately $N\sim10^6$ atoms in the c-field. This state is then phase-imprinted (as in the $T=0$ case) with a 2$\pi$ winding, and taken as the initial condition.

In Fig.~\ref{fig:4} we show example {\em single-trajectory} oscillations for ${T=50,150,250\text{nK}}$, from left to right, and fixed $\gamma=0.001$. Example initial states are shown on the top row of Fig.~\ref{fig:4}, with increasing fluctuations. Individual trajectories of the persistent current oscillations are shown in row (ii). Fluctuations reduce visibility of the oscillations with increasing temperature, essentially shifting the phase of the oscillation. This phase shift washes out the signal of the average winding number, $\langle n_{\text{L}}\rangle$, taken over 100 distinct numerical noise realizations in (iii). The absolute value of the average winding number 
is related to the probability of finding the current in the left, or right, ring.
We have chosen this quantity as the easiest to realize in an experiment, through repeated destructive measurements of the winding number. Even for relatively high temperatures, there are still clearly observable oscillations. In all cases, we find a final value $|\langle n_{\text{L}}\rangle|=0.5$, which corresponds to a random final configuration. 

Results from Fig.~\ref{fig:3} revealed a decreasing oscillation lifetime with increasing dissipation rate $\gamma$.
As expected, in the limit of low temperatures [$T=50$nK case] where the fluctuations are relatively small, the average $\langle \Delta L_z \rangle$ over stochastic realizations [Fig.~\ref{fig:4}(a)(iv)] exactly coincides with the dissipative result of the damped GPE [Eq.~\eqref{eqn:dGPE}] with the same dissipation parameter $\gamma$ [Fig.~\ref{fig:3}(a)(v)], the latter shown by the red curve in Fig.~\ref{fig:4}(a)(iv).
However, as evident from Fig.~\ref{fig:4}, stronger noise (corresponding to a higher temperature) may conceal some features, pointing towards an even shorter period over which oscillations are detectable in the presence of fluctuations, as expected to be relevant in realistic experimental settings.
%

Our analysis so far has been based on purely 2D simulations, even though the chosen parameter regime ($N=10^6$, $\mu_\text{2D} \sim 3.6 \hbar \omega_z$) does not automatically guarantee the 2D nature of our system, allowing for the potential of occupation of multiple transverse modes. Moreover, appealing as it may be due to the inclusion of stochastic fluctuations and mimicking actual experimental observables, the SPGPE model in fact gives us a cumulative measurement of parameters over the c-field region of highly-occupied modes.
If one is interested in more directly identifying the effect on the condensate mode, within a 3D setting which also fully accounts for transverse degrees of freedom, one could alternatively use a 3D kinetic model, to which we next turn our attention.

\begin{figure*}[ht]
    \centering
    \includegraphics[width=2\columnwidth]{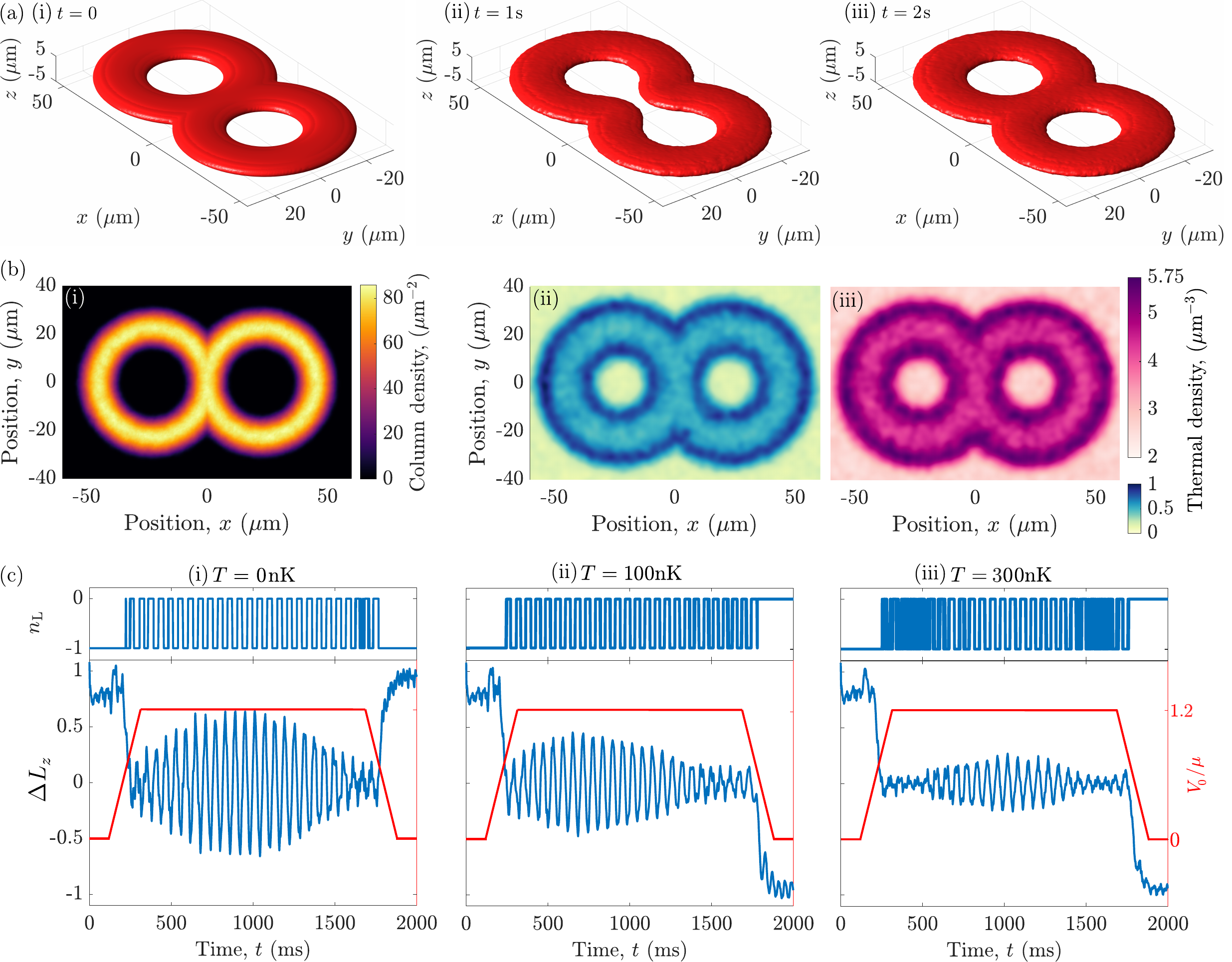}
    \caption{Persistent current oscillations in 3D. (a) condensate density isosurfaces at (i) $t=0$ (ii) $t=1$s and (iii) $t=2$s. Isosurface level is at 5\% of the maximum density. (b)(i) Initial condensate column density for $T=300$nK. Other temperatures show similar distributions. (b)(ii) and (iii) Thermal cloud density slices through $n(x,y,z=0)$ for (ii) $T=100$nK and (iii) $T=300$nK. (c) Example persistent current oscillation and angular momentum difference between rings for a single run, for increasing temperature with each column. Red curve indicates barrier ramp protocol.}
    \label{fig:5}
\end{figure*}

\section{Oscillations in a 3D kinetic model}\label{sec:3}

Finally, we extend our analysis to a fully three-dimensional (3D) finite temperature system, to fully and self-consistently numerically account for the effect of the perpendicular harmonic trap on the persistent current oscillations.
Rather than implementing the same SPGPE model in 3D, we instead choose to describe the system in terms of a kinetic model which facilitates a direct distinction between the condensate and the self-consistently coupled thermal cloud.

\subsection{Collisionless ZNG Model: Gross-Pitaevskii coupled to a Collisionless Boltzmann Equation}

At finite temperature, the bosonic quantum gas is partially condensed and we must consider the presence of the thermal cloud. We have used the collisionless version of the Zaremba-Nikuni-Griffin (ZNG) model~\cite{griffin2009bose} to describe the behavior of this system. In this model the condensate mode dynamics are described by a generalized GPE, which includes an additional term accounting for the mean field potential of the thermal cloud, $2gn_\text{th}$, which can be thought of as a time-dependent correction to the 3D trapping potential $V_\text{ext}({\bf \rho},z,t)$.
The condensate mode, $\Psi({\bf \rho},z,t)$, thus obeys the 3D equation~\cite{griffin2009bose}
\begin{align}
i\hbar \frac{\partial\Psi}{\partial t} = \left(-\frac{\hbar^2}{2m}\nabla^2+V_\text{ext}+g(|\Psi|^2 + 2n_\text{th})\right)\Psi\,,
\label{eqn:GGPE}
\end{align}
where the 3D scattering amplitude is $g=4\pi\hbar^2a_s/m$. 
This equation is solved self-consistently with a collisionless Boltzmann equation for the single--particle phase-space distribution, $f(\textbf{r},\textbf{p},t)$. The single--particle distribution function is defined as the number of particles within a neighborhood of the phase-space point $({\bf r},{\bf p})$ at time $t$. The thermal--cloud density, $n_{\rm th}({\bf r},t)$, at time $t$ can extracted from $f({\bf r},{\bf p},t)$ by integrating over ${\bf p}$:
\begin{equation}
n_{\rm th}({\bf r},t) = 
\int\,\text{d}^{3}{\bf p}\,f({\bf r},{\bf p},t).
\end{equation}
The collisionless Boltzmann equation for $f$ has the form
\begin{align}
    \frac{\partial f}{\partial t} + \frac{\textbf{p}}{m}\cdot\nabla_\textbf{r}f-\nabla_\textbf{r}V_\text{eff}\cdot\nabla_\textbf{p}f=0\,,
\end{align}
The effective potential felt by the thermal atoms now takes the form $V_\text{eff} = V_\text{ext}+2g(|\Psi|^2 + n_\text{th})$. The ZNG kinetic theory has been successfully used to model a range of dynamical phenomena in single- and multi-component condensates (see, e.g.~Refs.~\cite{griffin2009bose,Lee_Proukakis_2016,xhani2020critical,berloff_brachet_14,eller2020producing,xhani2021dissipation} and references therein).

The initial thermal-equilibrium wave function, $\Psi_0$, and thermal cloud density, $n_\text{th}^0$, are set by the temperature-dependent system chemical potential $\mu(T)$, with
the initial finite-temperature equilibrium distribution obtained iteratively for a fixed total atom number, as described in Ref.~\cite{griffin2009bose,jackson2002modeling}.
For consistency, and easier interpretation of our results, throughout our $T>0$ simulations we ensure that the BEC number is equal to the corresponding $T = 0$ number, fixed here to $10^6$ particles. 
As a result, the {\em total} particle number in the system increases with increasing temperature, and in this section we investigate the dissipative role of the thermal cloud, up to the point where the condensate and thermal atoms become comparable (i.e.~a $\sim$50\% condensate fraction). 

To ensure a comparable final barrier height across all $T>0$ simulations, throughout our present analysis the barrier height is fixed in terms of the temperature-dependent chemical potential according to $V_0 = 1.2 \mu(T)$.
As before, a 2$\pi$ winding is imprinted in the left ring 100ms before $t=0$.  An initial condensate density isosurface is shown in Fig.~\ref{fig:5}(a)(i), and corresponding column density in (b)(i).

Firstly, we confirm the persistence of undamped current oscillations in a $T=0$ 3D system. 
As before, we restrict our dynamical simulations to a barrier opening (approximately) consistent with a beat half-cycle, and report such undamped oscillations in Fig.~\ref{fig:5}(c)(i).
The oscillation period is found to be $\sim$60ms, slightly longer than the 2D model but still within the analytic prediction, which is still valid assuming the vortex traverses along $z=0$ and $\hat{\bm{\kappa}} = \kappa s \cdot \hat{\textbf{e}}_{z}$. 
Similar oscillations are observed in $\Delta L_z$, with evidence of the beating effect between persistent current and atom number oscillations clearly visible also here
\footnote{Use of a slightly different initialization condition in 3D leads to the observed beating starting at a different point in its cycle, such that--when the barrier is opened up--the $\Delta L_z$ oscillations about zero initially increase in amplitude, before decreasing again to reach the beat half cycle; this does not however affect our results in any way, and different offsets are also seen in the subsequent $T>0$ simulations.}.
Finally, we note that, although we only show results for $V_0=1.2\mu_{\text{2D}}(T)$ here, oscillations are found already when the maximum barrier amplitudes are as low as $V_0=0.8\mu_{\text{2D}}(T)$.

At finite temperatures, the thermal cloud tends to build up in the low density regions surrounding the condensate, as shown in the thermal equilibrium density slices at $z=0$ in Fig.~\ref{fig:5}(b) for $T=100$nK [panel (ii)] and $T=300$nK [panel (iii)] corresponding to the condensate density of panel (i). 
The oscillations are still clearly visible in both $n_\text{L}$ and $\Delta L_z$ for $T=100$nK, but (as already mentioned in the context of the damped GPE) the low amplitude oscillations in $\Delta L_z$ can wash out the visibility of the persistent current oscillations at higher temperature ($T=300$nK). 
We also see here that the maximum amplitude of the $\Delta L_z$ beat decreases with increasing temperature. This is a clear signature of damping due to the dynamical coupling of the condensate to the thermal cloud. Nonetheless, the key underlying feature of $\Delta L_z$ oscillations about a zero value remains detectable even at non-negligible finite temperatures (the $T=300$nK case has a near $\sim50$\% depletion), with $|\Delta L_z|<0.5$ across all probed 3D regimes.


Our results here, corresponding to single experimental runs, are visually much cleaner than the SPGPE c-field analysis. This is due to the direct access to the condensate mode facilitated within the present model. Importantly, however, the 3D nature of the simulations completely replicates the features analyzed in detail in 2D in earlier sections.

\section{Conclusions}

We have theoretically demonstrated the ability to control the periodic transfer of persistent current across two rings in which a quasi-2D quantum gas is trapped.
Simulations for a pure atomic condensate have clearly confirmed the stability of oscillations of the state between the two rings. Our extensive 2D and 3D simulations conducted in the context of a quasi-2D geometry, have further revealed such oscillations to be long-lived, even at finite temperature, based on two distinct state-of-the-art finite temperatures models. At low temperatures and with minimal damping, these oscillations dissipate until the vortex, the carrier of the persistent current, sits in the center of the system. If the damping is large enough, there are no oscillations.

These results were backed up by an analytic model for the vortex dynamics, assuming the vortex traverses the low density region on the central edge of the rings. This same model qualitatively predicts the critical damping parameter at which the oscillations are halted. 
Our findings should be within observational reach based on current experimental capabilities and detection schemes (see, e.g.~Refs.~\cite{amico2021roadmap, eckel2014hysteresis, kumar2016minimally,kumar15-detection, corman2014quench, aidelsburger2017relaxation,aghamalyan2015coherent,amico_readout,Eckel2014-PRX}) and pave the way for future quantum technological devices and sensors.
For example, we envisage our model will be applicable for precise measurements of rotation and acceleration. Under large damping, the vortex is known to sit at the center of the system, however external rotation or acceleration will affect the vortex's final position. Applications of our work to an accelerometer will be the subject of future work.

Data supporting this publication are openly available under a Creative Commons CC-BY-4.0 License in Ref\,\cite{data}.

\section{Acknowledgements}

We gratefully acknowledge Oksana Chelpanova for stimulating discussions. TB acknowledges an EPSRC Doctoral Prize Fellowship (Grant No. EP/R51309X/1) and a joint-project grant from the FWF (Grant No. I4426, RSF/Russland 2019). 
TB and NPP also acknowledge support from the Quantera ERA-NET cofund project NAQUAS through the Engineering and Physical Science Research Council (Grant No.EP/R043434/1). AY acknowledges the National Research Foundation of Ukraine (2020.02/0032).
ME acknowledges support from the US National Science Foundation grant no.\,PHY-1707776. Such collaboration was initiated within the framework of the Benasque Atomtronics Workshops, which we gratefully acknowledge.


\begin{thebibliography}{58}
\bibitem [{dat()}]{data}%
  \BibitemOpen
  \href@noop {} {\bibinfo {title} {{Newcastle University Data (data.ncl.ac.uk)
  doi: to be added.}}}\BibitemShut {Stop}%
\end{thebibliography}
\end{document}